\documentclass[useAMS,usenatbib, parskip=no]{mn2e}
\usepackage{float}
\usepackage{graphicx}

\title[Cluster Mass Function of Young Clusters]
    {On the origin of the Schechter-like mass function\\ of young star clusters in disk galaxies}
\author[P. Lieberz et al.]
  {P.~Lieberz$^1$,
  P.~Kroupa$^{1,2}$
  \newauthor % starts a new line in the
\\
  $^1$ Helmholtz Institut f\"ur Strahlen- und Kernphysik (HISKP), Universit\"at Bonn, Nussallee 14--16, D-53121 Bonn, Germany\\
 $^2$ Charles University in Prague, Faculty of Mathematics and Physics, Astronomical Institute,\\ \hspace{0.1cm} V  Hole\v{s}ovi\v{c}k\'ach 2, CZ-180 00 Praha 8, Czech Republic}

\date{Released 2002 Xxxxx XX}

\pagerange{\pageref{firstpage}--\pageref{lastpage}} \pubyear{2002}

\def\LaTeX{L\kern-.36em\raise.3ex\hbox{a}\kern-.15em
    T\kern-.1667em\lower.7ex\hbox{E}\kern-.125emX}

\begin{document}

\label{firstpage}

\maketitle

\begin{abstract}
The mass function of freshly formed star clusters is empirically often described as a power law. However the cluster mass function of populations of young clusters over the scale of a galaxy has been found to be described by a Schechter-function. Here we address this apparent discrepancy.
We assume that in an annulus of an isolated self-regulated radially-exponential axially-symmetric disk galaxy, the local mass function of very young (embedded) clusters is a power law with an upper mass limit which depends on the local star formation rate density.
Radial integration of this mass function yields a galaxy-wide embedded cluster mass function.
This integrated embedded cluster mass function has a Schechter-type form, which results from the addition of many low mass clusters forming at all galactocentric distances and rarer massive clusters only forming close to the center of the galaxy.
\end{abstract}


\section{Introduction}

The freshly-formed stellar mass of a galaxy resides in its embedded star clusters (\citealt{ll}; \citealt{pakr}; \citealt{meg}; \citealt{mei}), which can synonymously be referred to as (about 1 pc extended, 1 Myr duration) space-time correlated star-formation events. i.e. essentially in molecular cloud clumps. The investigation of embedded clusters helps us to understand the build up of the stellar populations in galaxies.
Embedded clusters are still fully or partially enshrouded in gas and dust, and represent the earliest stage in the life-time of a formed star cluster. They may be the forerunners to the open clusters and are therefore valuable probes for cluster formation \citep{ll}.\par
Observations indicate that the masses of the embedded star clusters in galaxies follow a particular distribution function, the embedded cluster mass function (ECMF). Locally it is typically found to be a power-law, while galaxy-wide observation reveal a Schechter-like turn-down \citep{gl}. This has also been observed for old, massive globular clusters (see e.g. \citealt{jord}; \citealt{bur}; \citealt{par}). Massive clusters experience limited secular mass loss and are therefore still a reasonable indicator for the initial cluster mass function \citep{baum}.\par
The focus of this paper is on embedded clusters in the disks of isolated late-type galaxies. Ideally the disk can be seen as a purely self-regulated axis-symmetric system, such that it is possible to investigate the ECMF only radially over the area of the disk galaxy to explain the difference between the local and integrated distribution\footnote{The assumption of the galaxy being axis-symmetric is made for computational ease. The model developed here is equally valid in any galaxy in which the most mast massive cluster forming in a population of embedded clusters depends on the local gas density.}.\par
In this paper we present for the first time analytical approaches to the galaxy wide ECMF and also rewrite the dependency on the galactocentric distance into a mass dependency so that we arrive at a new relation between the number and the mass of star clusters. Hereby we follow the ansatz that the galaxy-wide ECMF is the sum of all local ECMFs.\par
We constrain these models with observational data and show that the models account for the data quite well.

\section{The local embedded cluster mass function}
\label{LECMF}
The local ECMF (LECMF or $\xi_{\rmn{lecl}}$) describes the surface number density of star clusters in the stellar mass interval \mbox{$\left[ M_{\rmn{ecl}}, \, M_{\rmn{ecl}}+\mathrm{d}M_{\rmn{ecl}} \right]$} in an infinitesimally small surface area ($\mathrm{d}A$) at a distance $r$ from the center of the star-forming disk galaxy
\begin{equation}
	\label{definition}
	\xi_{\rmn{lecl}}(M_{\rmn{ecl}}; r)\,\mathrm{d}M_{\rmn{ecl}} \, \mathrm{d}A=\mathrm{d}N_{\rmn{ecl}}\, ,
\end{equation}
where $\mathrm{d}N_{\rmn{ecl}}$ is the number of embedded clusters in $\mathrm{d}M_{\rmn{ecl}}$ and $\mathrm{d}A$.\par
In general the LECMF has the form of a power law, as derived from observations \citep{ll}

\begin{equation}
	\label{LECMF}
	\xi_{\rmn{lecl}}(M_{\rmn{ecl}}; r) = K(r) M_{\rmn{ecl}}^{-\beta},
\end{equation}
where $M_{\rmn{ecl}}$ is the stellar mass of the embedded cluster at ``birth''\footnote{``birth'' describes here the idealised state when the entire population is existing directly prior to the expulsion of the gas.} and $K(r)$ the normalization constant and $\beta$ the power law index.
By observation $\beta$ was found to be between $\beta=1.5$ and $\beta=2.5$ (\citealt*{wkl}, hereafter WKL, and references therein). $K(r)$
has to be estimated using normalization conditions.\par
The normalization condition is that the entire freshly formed stellar mass of a galactic region has to be in embedded clusters. So integrating the mass over all clusters (from the lower mass limit $M_{\rmn{ecl,min}}$ to the locally upper mass limit $M_{\rmn{U,loc}}(r)$) obtains the freshly formed stellar mass. In the case of an infinitesimally small surface area $\mathrm{d}A$ this obtains the embedded cluster mass surface density. This embedded cluster mass surface density is defined as the star formation rate density $\Sigma_{\rmn{SFR}}(r)$ multiplied with the time-scale ,$\delta t$, which is the time over which a population of embedded clusters forms.
$\Sigma_{\rmn{SFR}}(r)$ is defined as

\begin{equation}
	\label{SSFR}
	\Sigma_{\rmn{SFR}}(r) = \frac {\mathrm{d}\rmn{SFR}}{\mathrm{d}A} \, ,
\end{equation}
with $\rmn{SFR}$ being the star formation rate in the whole galaxy.
The total embedded cluster mass density is obtained by multiplying $\Sigma_{\rmn{SFR}}(r)$ with $\delta t$ (\citealt*{sch}; WKL):
\begin{equation}
	\label{Correct}
	\Sigma_{\rmn{SFR}}(r) \, \delta t= \int_{M_{\rmn{ecl,min}}}^{M_{\rmn{U,loc}}(r)} M_{\rmn{ecl}}^{\prime}\xi_{\rmn{lecl}}(M_{\rmn{ecl}}^{\prime};r)\mathrm{d}M_{\rmn{ecl}}^{\prime}\, .
\end{equation}
$M_{\rmn{ecl,min}}$ is assumed to be about 5 $M_{\odot}$, corresponding to the smallest groups of embedded stars observed (\citealt{kb}, \citealt{km}), and is here assumed to be a constant. $\delta t$ is roughly 10 Myr, as deduced by \citet*{egu}. This is the time it takes for the interstellar medium to transform into a new population of stars and is essentially the lifetime of molecular clouds (\citealt{fuk}; \citealt{yam}; \citealt{tam}). It corresponds to the lifetime over which the embedded cluster mass function is fully populated (see the discussion in \citealt{pa} and \citealt{sch}).\par
\begin{figure}
    \begin{center}
            \resizebox{0.5\textwidth}{!}{\input{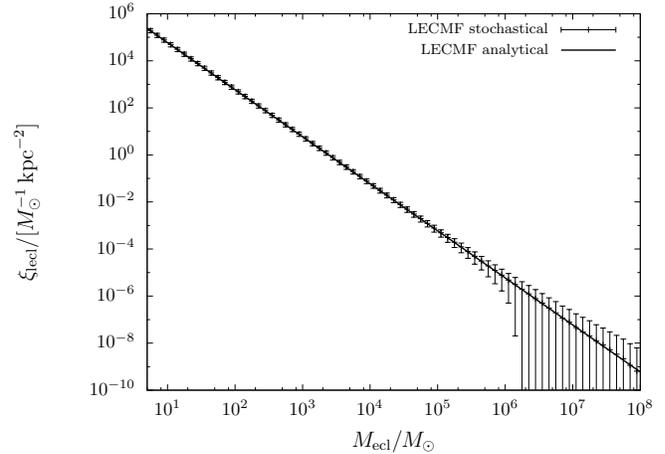}}
    \end{center}
    \caption{
    An exemplary LECMF. Shown here is a LECMF from the distribution function given by Eq. (\ref{LECMF}) for a galactic region with $10^8$ $M_{\odot}$ of stellar mass. Further we assumed a $M_{\rmn{ecl,min}}$ of 5 $M_{\odot}$ and a $M_{\rmn{U,loc}}(r)$ of $10^8$ $M_{\odot}$. The solid line shows the LECMF calculated analytically, whereas the points are the binned results of stochastic sampling.
		}
   \label{LECMFfig}
\end{figure}

It is noteworthy that to calculate the LECMF Eq. (\ref{Correct}) is sufficient once $M_{\rmn{U,loc}}(r)$ is known. The values for $K(r)$ are uniquely defined for different $\beta$. For $\beta \ne 2$,

\begin{equation}
	\label{Kbn2}
	K(r) = \frac{\Sigma_{\rmn{SFR}}(r) \, \delta t (2-\beta)}{M_{\rmn{U,loc}}^{2-\beta}(r)-M_{\rmn{ecl,min}}^{2-\beta}}\, .
\end{equation}
For the special case of $\beta =2$,

\begin{equation}
	\label{Kb2}
	K(r) = \frac{\Sigma_{\rmn{SFR}}(r) \, \delta t }{\ln \left( M_{\rmn{U,loc}}(r)/M_{\rmn{ecl,min}} \right) }\, .
\end{equation}
A visual verification that a LECMF calculated by this method is in agreement with a LECMF that was obtained by randomly drawing star clusters according to Eq. (\ref{LECMF}) is shown in Fig. \ref{LECMFfig}.\par

The only uncertainty left is the exact form of $M_{\rmn{U,loc}}(r)$.
At the moment we can only state that the local upper mass limit has to be smaller than the cluster upper mass limit of the whole galaxy $M_{\rmn{U}}$:
$M_{\rmn{U,loc}}(r) \le M_{\rmn{U}}$.
Therefore additional research into this subject is necessary (see Sec. \ref{emj}).\par
The LECMF with an observable non-infinitesimal surface area $\Delta A$ is discussed in Appendix \ref{DeltaA}.

\section{The Galaxy-Wide Embedded Cluster Mass Function}
\label{IECMF}
In this section the scope of our analysis is to describe the form of the ECMF after integrating over the axis-symmetric galactic disk surface area.
The previous formulae allow the normalisation of the LECMF at a fixed distance to the center of a galaxy.\par
The galaxy wide or integrated ECMF (IECMF or $\xi_{\rmn{iecl}}$, not to be confused with the LECMF $\xi_{\rmn{lecl}}$ above) is defined as the number of star clusters in the stellar mass interval $M_{\rmn{ecl}}$ to \mbox{$M_{\rmn{ecl}}+\mathrm{d}M_{\rmn{ecl}}$}
\begin{equation}
	\label{definition2}
	\xi_{\rmn{iecl}} (M_{\rmn{ecl}})\, \mathrm{d}M_{\rmn{ecl}}=\mathrm{d}N_{\rmn{ecl}}\, ,
\end{equation}
or
\begin{equation}
	\label{definition3}
	\begin{array}{r r l}
	\xi_{\rmn{iecl}} (M_{\rmn{ecl}})&=&\int_{0}^{2 \pi} \int_{0}^{R^{\prime}(M_{\rmn{ecl}})}\xi_{\rmn{lecl}}(M_{\rmn{ecl}};r)\,r \, \mathrm{d}r \mathrm{d}\phi \vspace{2mm} \\
	&=&\int_{0}^{2 \pi} \int_{0}^{R^{\prime}(M_{\rmn{ecl}})} K(r) M_{\rmn{ecl}}^{-\beta}r\,\mathrm{d}r \mathrm{d}\phi \, .
	\end{array}
\end{equation}
with $R^{\prime}(M_{\rmn{ecl}})$ being the maximal galactocentric distance at which an embedded cluster of mass $M_{\rmn{ecl}}$ can form, assuming the maximum cluster mass decreases monotonically with increasing radial distance. \par
Is the IECMF still similar to a power-law function of the cluster mass? And if yes, does the power-law parameter $\beta$ differ from the local one? These questions are addressed in the following.\par 
The galaxy wide star formation rate (SFR)
\begin{equation}
	\label{defSFR}
	\rmn{SFR}=\int_{0}^{2 \pi} \int_{0}^{R_{\rmn{gal}}}\Sigma_{\rmn{SFR}}\,r \, \mathrm{d}r \mathrm{d}\phi\, ,
\end{equation}
with $R_{\rmn{gal}}$ being the radius of the star forming area of the the galaxy.
In order to obtain $\xi_{\rmn{iecl}}$ one can insert Eq. (\ref{Correct}) into Eq. (\ref{defSFR}).
The resulting equation (see also \citealt{pa}) is:
\begin{equation}
	\label{Globalold}
		\begin{array}{r r l}
	\rmn{SFR} \, \delta t&=& \int_{0}^{2 \pi} \int_{0}^{R_{\rmn{gal}}} \int_{M_{\rmn{ecl,min}}}^{M_{\rmn{U,loc}}(r)} M_{\rmn{ecl}}\xi_{\rmn{lecl}}(M_{\rmn{ecl}};r)\mathrm{d}M_{\rmn{ecl}} r \mathrm{d}r \mathrm{d}\phi \vspace{2mm} \\
	&=& \int_{0}^{2 \pi} \int_{0}^{R_{\rmn{gal}}} \int_{M_{\rmn{ecl,min}}}^{M_{\rmn{U,loc}}(r)} K(r) M_{\rmn{ecl}}^{1-\beta} r\, \mathrm{d}M_{\rmn{ecl}} \mathrm{d}r \mathrm{d}\phi\, .
	\end{array}
\end{equation}
Eq. (\ref{Globalold}) is first an integration over the mass of the clusters and then over the area.
It is possible to exchange the mass integral with the integral over the galactocentric distance by noting that in a given annulus the theoretical most massive cluster $M_{\rmn{U,loc}}(r)$ depends on $r$, that is, we can invert this function to obtain the galactocentric distance, $R^{\prime}(M_{\rmn{ecl}})$, at which the annulus contains a particular theoretical most massive cluster.
In other words to exchange the positions of the $r$- and $M_{\rmn{ecl}}$-integration.
Assuming that the mass of the theoretical most massive cluster possible, $M_{\rmn{U,loc}}(r)$, decreases with the galactocentric distance $r$ then the integration needs only to extend over the distances 0 to \mbox{$R^{\prime}(M_{\rmn{ecl}}) \le R_{\rmn{gal}}$}. Therewith at the low mass end the IECMF is an integral over 0 to $R_{\rmn{gal}}$, while at the high mass end large radii do not contribute:
\begin{equation}
	\label{Global}
	\rmn{SFR} \, \delta t = \int_{M_{\rmn{ecl,min}}}^{M_{\rmn{U}}} \int_{0}^{2 \pi}  \int_{0}^{R^{\prime}(M_{\rmn{ecl}})}  K(r) M_{\rmn{ecl}}^{1-\beta} r\, \mathrm{d}r \mathrm{d}\phi\mathrm{d}M_{\rmn{ecl}}\, .
\end{equation}
Neither $K(r)$ nor $M_{\rmn{U,loc}}(r)$ are known for a specific $r$. $K(r)$ depends on $\Sigma_{\rmn{SFR}}$, see Eq. (\ref{Correct}).
A direct relation between $\Sigma_{\rmn{SFR}}$ and the position within the galaxy, assumed to be valid for a galaxy in self-regulated equilibrium, is taken from \citet{jp}:

\begin{equation}
	\label{sigmasfr}
	\begin{array}{r r l}
		\Sigma_{\rmn{SFR}}(r)&=&\frac{\rmn{SFR} \, \,\, e^{R_{\rmn{gal}}/r_{d}}}{2\pi r_{d}^2 \left( e^{R_{\rmn{gal}}/r_{d}} - \frac{R_{\rmn{gal}}}{r_{d}} - 1 \right)}e^{-r/r_{d}} \vspace{2mm} \\
		&\approx&\frac{\rmn{SFR}}{2\pi r_{d}^2}e^{-r/r_{d}} \, .
	\end{array}
\end{equation}
In this case $r_{d}$ is the disc scale length. The equation is normalised in such a way that an integral over the whole area results in the total SFR. Therefore both sides of Eq. (\ref{sigmasfr}) can be multiplied with $\delta t$ so that it is equal to Eq. (\ref{Correct}) (in the following we will write \mbox{SFR $\delta t$} as $M_{\rmn{tot}}$, the total stellar mass formed galaxy wide in time $\delta t$):

\begin{equation}
	\label{getKMU}
	\int_{M_{\rmn{ecl,min}}}^{M_{\rmn{U,loc}}(r)} M_{\rmn{ecl}}^{\prime}\xi_{\rmn{lecl}}(M_{\rmn{ecl}}^{\prime})\mathrm{d}M_{\rmn{ecl}}^{\prime}=\frac{M_{\rmn{tot}}}{2\pi r_{d}^2}e^{-r/r_{d}} \, .
\end{equation}
This gives a relation between $K(r)$ and $M_{\rmn{U,loc}}(r)$.  For $\beta \neq 2$:
\begin{equation}
	\label{getKMU2}
	K (r) = \frac{\left( 2-\beta \right) M_{\rmn{tot}}}{2\pi r_{d}^2 \left(M_{\rmn{U,loc}}^{2-\beta}(r)-M_{\rmn{ecl,min}}^{2-\beta} \right)}e^{-r/r_{d}} \, .
\end{equation}
For $\beta=2$:
\begin{equation}
	\label{getKMU3}
	K (r) = \frac{ M_{\rmn{tot}}}{2\pi r_{d}^2 \ln \left( M_{\rmn{U,loc}}(r)/M_{\rmn{ecl,min}} \right) }e^{-r/r_{d}}\, .
\end{equation}

To get an unambiguous expression for $K(r)$ and $M_{\rmn{U,loc}}(r)$ more constraints are needed. For this purpose we use a model (henceforth called the exponential model) based on an ansatz from \citet{jp}. Other possible models, found to be not working as well as this one, are discussed in Appendix \ref{OTHER}.

\subsection{Exponential Model}
\label{emj}
\citet{jp} propose the ansatz that the radial dependence of $M_{\rmn{ecl,max,loc}}(r)$ should have the same form as the radial dependence of the gas surface density $\Sigma_{\rmn{gas}}(r)$:
\begin{equation}
	\label{Jangas}
	\Sigma_{\rmn{gas}}(r) = \Sigma_{\rmn{gas,0}}\, e^{-\frac{r}{r_d}}\, ,
\end{equation}
with $\Sigma_{\rmn{gas,0}}$ being the gas surface density at the center of the galactic disk. Thus
\begin{equation}
	\label{Janold}
	M_{\rmn{ecl,max,loc}}(r) = M_{\rmn{ecl,max}}\, e^{-\frac{r}{r_d}}\, .
\end{equation}
With this ansatz they were able to show that the $\rmn{H}\alpha$ radial cut-off in disk galaxies is naturally explained, given that star formation extends well beyond this cut-off radius.\par
As our model uses $M_{\rmn{U,loc}}$ instead of \mbox{$M_{\rmn{ecl,max}} $}, we modify Eq. (\ref{Janold}) to
\begin{equation}
	\label{Jannew}
	M_{\rmn{U,loc}}(r) = M_{\rmn{U}} e^{-\frac{r}{r_d}}\, .
\end{equation}
Inserting Eq. (\ref{Jannew}) into Eq. (\ref{getKMU}) with $\beta \neq 2$ results in a definite $K(r)$:
\begin{equation}
	\label{JanKnb}
	K (r) = \frac{\left( 2-\beta \right) M_{\rmn{tot}}}{2\pi r_{d}^2 \left[\left( M_{\rmn{U}} e^{-\frac{r}{r_d}} \right)^{2-\beta}-M_{\rmn{ecl,min}}^{2-\beta} \right]}e^{-r/r_{d}} \, ,
\end{equation}
and for $\beta = 2$ in
\begin{equation}
	\label{JanKb}
	K (r) = \frac{ M_{\rmn{tot}}}{2\pi r_{d}^2 \ln \left[ \left( M_{\rmn{U}}e^{-\frac{r}{r_d}} \right)/M_{\rmn{ecl,min}} \right] }e^{-r/r_{d}}\, .
\end{equation}
For each $r$ in the axis-symmetric disk galaxy there is a theoretical maximal cluster mass $M_{\rmn{U,loc}}(r)$. Regarding the entire galaxy, each $M_{\rmn{U,loc}}(r)$ is a theoretical possible cluster mass $M_{\rmn{ecl}}$. And in this relation $r$  is the maximal galactocentric distance $R^{\prime}(M_{\rmn{ecl}})$, at which a cluster of mass $M_{\rmn{ecl}}$ can still be found. This is true for every $M_{\rmn{ecl}}$, $R^{\prime}(M_{\rmn{ecl}})$ being the reverse function of $M_{\rmn{U,loc}}(r)$. $R^{\prime}(M_{\rmn{ecl}})$ is needed for Eq. (\ref{Global}),\par
\begin{equation}
	\label{JanR}
	R^{\prime}(M_{\rmn{ecl}}) = -r_{d}\ln \left(\frac{M_{\rmn{ecl}}}{M_{\rmn{U}}} \right)\, .
\end{equation}
Now $\xi_{\rmn{iecl}} (M_{\rmn{ecl}})$ can be calculated:
\begin{equation}
	\label{Janfinal}
	\xi_{\rmn{iecl}} (M_{\rmn{ecl}})=\int_{0}^{2 \pi} \int_{0}^{R^{\prime}(M_{\rmn{ecl}})}K(r) M_{\rmn{ecl}}^{-\beta}\,r \, \mathrm{d}r \mathrm{d}\phi\, .
\end{equation}
This integration can only be solved numerically (because of the $r$-dependence in $K(r)$).
The remaining free parameters ($\beta$ and $M_{\rmn{U}}$) can be fixed using empirical data. WKL derived a fitting function for the mass of the most massive very young cluster in a galaxy ($M_{\rmn{vyc,max}}$) depending on the SFR of the host galaxy:
\begin{equation}
	\label{maxfit2}
	M_{\rmn{vyc,max}} = k_{\rmn{ML}} \cdot \rmn{SFR}^{0.75(\pm 0.03)} \cdot 10^{6.77(\pm0.02)}  ,
\end{equation}
with SFR in $M_{\odot}/\rmn{yr}$ and $k_{\rmn{ML}}$ being the mass-to-light ratio in the photometric V-band, which can be assumed to be 0.0144 $M_{\odot}/L_{V,\odot}$ for young (\mbox{$< 6$ Myr}) clusters. $M_{\rmn{vyc,max}}$ is a good approximation for the stellar mass of the most massive embedded star cluster of a galaxy, $M_{\rmn{ecl,max}}$.\par 
To determine $M_{\rmn{ecl,max}}$ we use two conditions: first there is only one most massive cluster. Second the mass of the most massive cluster is $M_{\rmn{ecl,max}}$.\par
To implement the first condition we choose a mass interval between the upper mass limit $M_{\rmn{U}}$ and $M_{\rmn{ecl,t}}$, with 
$M_{\rmn{ecl,t}}$ chosen in such a way that there is only one cluster between these limits:
\begin{equation}
	\label{Irr1gJ}
	1 = \int_{M_{\rmn{ecl,t}}}^{M_{\rmn{U}}} \xi_{\rmn{iecl}}(M_{\rmn{ecl}}^{\prime})\mathrm{d}M_{\rmn{ecl}}^{\prime}\, .
\end{equation}
The second condition implies that the mass between these limits is 
$M_{\rmn{ecl,max}}$:
\begin{equation}
	\label{Irr2gJ}
	M_{\rmn{ecl,max}} = \int_{M_{\rmn{ecl,t}}}^{M_{\rmn{U}}} M_{\rmn{ecl}}^{\prime} \xi_{\rmn{iecl}}(M_{\rmn{ecl}}^{\prime})\mathrm{d}M_{\rmn{ecl}}^{\prime}\, .
\end{equation}
As $M_{\rmn{ecl,max}}$ depends on SFR it therewith follows that also $M_{\rmn{U}}$ depends on SFR. Note that these conditions differ from those employed on previous occasions, as explained in Appendix \ref{Weidnernorm}.\par
\begin{figure}
     \begin{center}
            \resizebox{0.5\textwidth}{!}{\input{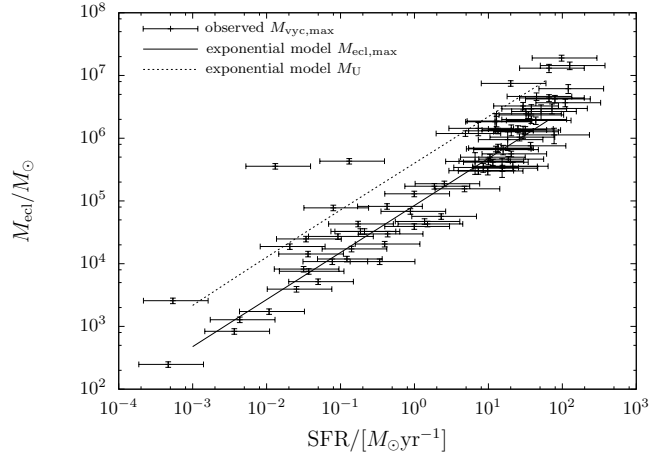}}
    \end{center}
    \caption[The observational data was taken from WKL for $M_{\rmn{vyc,max}}<2\cdot 10^{5} \, M_{\odot}$  and \citet{rev} for  $M_{\rmn{vyc,max}}>2\cdot 10^{5} \, M_{\odot}$ and shows the observed galaxy-wide most massive very young clusters $M_{\rmn{vyc,max}}$ (which are a good approximation for $M_{\rmn{ecl,max}}$) in dependence of SFR. Also plotted here is $M_{\rmn{ecl,max}}$  and $M_{\rmn{U}}$, as calculated using the model from Sec. \ref{emj} for $\beta=2.3$ against the galaxy-wide SFR. All observed clusters should be below the $M_{\rmn{U}}$-line, which is mostly the case given the uncertainties.]{
    The observational data was taken from WKL for $M_{\rmn{vyc,max}}<2\cdot 10^{5} \, M_{\odot}$  and \citet{rev}\footnotemark[3] for  $M_{\rmn{vyc,max}}>2\cdot 10^{5} \, M_{\odot}$ and shows the observed galaxy-wide most massive very young clusters $M_{\rmn{vyc,max}}$ (which are a good approximation for $M_{\rmn{ecl,max}}$) in dependence of SFR. Also plotted here is $M_{\rmn{ecl,max}}$  and $M_{\rmn{U}}$, as calculated using the model from Sec. \ref{emj} for $\beta=2.3$ against the galaxy-wide SFR. All observed clusters should be below the $M_{\rmn{U}}$-line, which is mostly the case given the uncertainties.
		}
   \label{SFRMmaxJ}
\end{figure}

Using these we can find for every $\beta$ a SFR-$M_{\rmn{ecl,max}}$-curve that aligns with Eq. (\ref{maxfit2}). In order to narrow $\beta$ down, a fitting $M_{\rmn{U}}$ is needed. $M_{\rmn{U}}$ is supposed to be larger than any observed cluster mass but not so large that there would be unrealistic gaps between the mass of the most massive clusters and $M_{\rmn{U}}$. In the following we assume $r_d=2.15$ kpc, as this is roughly the determined disk scale length of the Milky Way (\citealt{bovy}; \citealt{Por}). With that we find for $\beta=2.3\pm0.1$ and $\delta t = 10$ Myr a SFR-$M_{\rmn{ecl,max}}$-curve that aligns with Eq. (\ref{maxfit2}) and resulting $M_{\rmn{U}}$ that fulfils the above criteria (see also Fig. \ref{SFRMmaxJ}). This way we can give for every SFR a corresponding $M_{\rmn{U}}$. E.g. for SFR $=1$ $M_{\odot}/\rmn{yr}$ we obtain $M_{\rmn{U}}=400000$ $M_{\odot}$.

\footnotetext[3]{In order to calculate $M_{\rmn{vyc,max}}$ from the observed absolute magnitude ($M_{\rmn{V}}$) the following formula has been used (Abdullah private communication):

\[
  M_{\rmn{V}} = 4.79 - 2.5 \log_{10}\frac{M_{\rmn{ecl,max}}}{k_{\rmn{ML}}}\, ,
\]
with $k_{\rmn{ML}}$ being the mass to light ratio.}

\section{Comparison to Empirical Data}
We already used empirical data in Sec. \ref{emj} to align the SFR-$M_{\rmn{ecl,max}}$-curve of the exponential model with the empirical fit by WKL (Eq. \ref{maxfit2}) and in doing so constrained the free parameters ($\beta$ and $M_{\rmn{U}}$) of the model. In the following we want to determine whether the model, using said constraints, is also in reasonable agreement with other observations. \par
\begin{figure}
    \begin{center}
            \resizebox{0.5\textwidth}{!}{\input{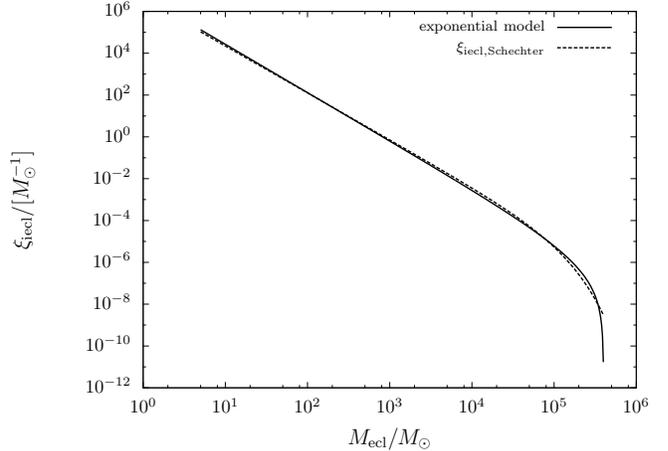}}
    \end{center}
    \caption{
    A comparison between the IECMF acquired from the exponential model (the solid line) and the IECMF from the Schechter form (Eq. \ref{Schechter}, the dashed line) for $\rmn{SFR}=1$ $M_{\odot}/\rmn{yr}$ and a $\delta t=10$ Myr. For the exponential model $\beta=2.3$, while for the $\xi_{\rmn{iecl, Schechter}}$ model $\beta=2.24$.
		}
   \label{ECMFS}
\end{figure}

Empirical data has been indicating that the galaxy-wide ECMF should be a Schechter-function \citep{gl}, i.e.
\begin{equation}
	\label{Schechter}
	\xi_{\rmn{iecl, Schechter}}(M_{\rmn{ecl}})=K^{\prime} e^{-M_{\rmn{ecl}}/M_{\rmn{c}}} M_{\rmn{ecl}}^{-\beta^{\prime}}\, .
\end{equation}
In this case $M_{\rmn{c}}$ is the turn-down mass, at which the ECMF turns down. Our formalism results in a similar form for the ECMF as the Schechter form. This is shown in  Fig. \ref{ECMFS}.\par
Note that the $\xi_{\rmn{iecl, Schechter}}$ model uses a different $\beta$ than the exponential model. The exponential model has a sharper turn down than the Schechter formalism for the same $\beta$. Therefore $\beta$ needs to be modified for a good fit. The smaller the SFR the bigger the divergence between the needed $\beta$ form the exponential model and the Schechter form.\par
Up until now no theoretical formulation existed which allowed $M_{\rmn{c}}$ to be predicted from properties of the galaxy. With our formulation we can derive the turn-down mass in dependence of the exponential density profile of the galaxy.\par
Fitting the Schechter-function to the exponential model allows us to determine a relation between $M_{\rmn{c}}$ and the SFR:
\begin{equation}
	\label{Mc_SFR}
	M_{\rmn{c}} = (85000 \pm 5000)\, \, \rmn{SFR}^{\left( 0.73 \pm 0.02 \right)}\, .
\end{equation}
Comparing this to Eq. (\ref{maxfit2}) shows that these two equations are the same within the uncertainties. Therefore $M_{\rmn{c}}=M_{\rmn{ecl,max}}$ is at least a good approximation.\par
\begin{figure}
    \begin{center}
            \resizebox{0.5\textwidth}{!}{\input{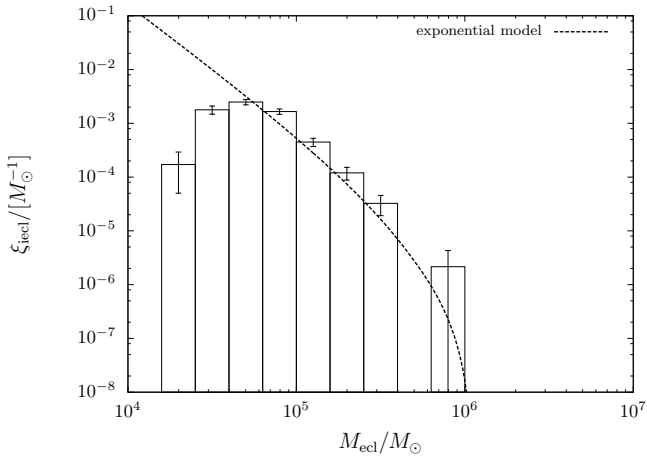}}
    \end{center}
    \caption{
     Combined binned young cluster mass function for two galaxies (NGC 5236 and NGC 6946, data taken from \citealt{lars}, \citealt{lars2} and Larsen private communication). Also shown is the exponential model. For the single galaxies see Figs. \ref{larsen5236} and \ref{larsen6946}.
		}
   \label{larsen}
\end{figure}

Next we compare the exponential model to actual empirical data. For this we confront the model with galaxy-wide observations of NGC 5236 and NGC 6946 (data taken from \citealt{lars}, \citealt{lars2} and Larsen private communication). The empirical data consists of a list of cluster masses, which were binned. To compare these data to the models we need the SFRs of the galaxies. \citet{dop} calculated a SFR of 2.76 $M_{\odot}/\rmn{yr}$ for NGC 5236, whereas \citet{hong} determined SFRs of 1.52 and 0.18 $M_{\odot}/\rmn{yr}$, depending on the method used. For NGC 6946 \citet{hee} measured, depending on the method, a SFR of \mbox{$4.6\pm 0.2$} and \mbox{$3.5\pm 0.2$} $M_{\odot}/\rmn{yr}$. For our models we are therefore adopting \mbox{$\rmn{SFR}=1.5$ $M_{\odot}/\rmn{yr}$} for NGC 5236 and \mbox{$\rmn{SFR}=4$ $M_{\odot}/\rmn{yr}$} for NGC 6946. From \citet{lars2} we get the ages of the brightest and 5th brightest cluster. From these we can assume that \mbox{$\delta t =10$ Myr} is a good approximation for the age. As \mbox{$\xi_{\rmn{lecl}} \propto K \propto \delta t$} a change in $\delta t$ does not change the overall behaviour of the function. A comparison of the data to the combined $\xi_{\rmn{iecl}}$ calculated using the exponential model is shown in Fig. \ref{larsen}.
\begin{figure}
    \begin{center}
            \resizebox{0.5\textwidth}{!}{\input{fig5.tex}}
    \end{center}
    \caption{
     Binned young cluster mass function for NGC 5236 (data taken from \citealt{lars}, \citealt{lars2} and Larsen private communication). Also shown is the exponential model for  \mbox{$\rmn{SFR}=1.5$ $M_{\odot}/\rmn{yr}$} and for \mbox{$\rmn{SFR}=2.76$ $M_{\odot}/\rmn{yr}$}. See text for further details.
		}
   \label{larsen5236}
\end{figure}
\begin{figure}
    \begin{center}
            \resizebox{0.5\textwidth}{!}{\input{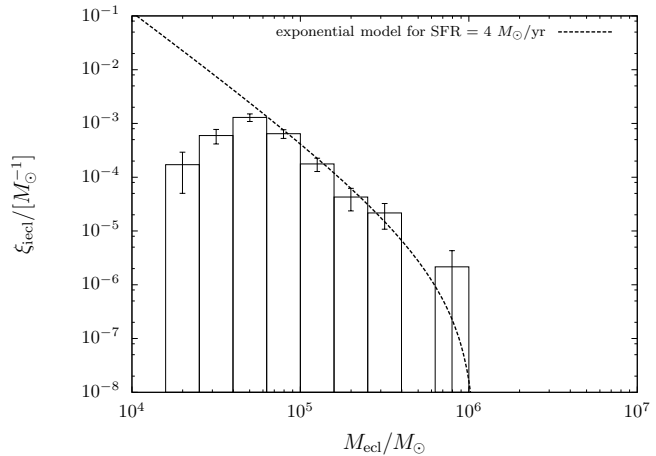}}
    \end{center}
    \caption{
    Binned young cluster mass function for NGC 6946 (data taken from \citealt{lars}, \citealt{lars2} and Larsen private communication). Also shown is the exponential model for \mbox{$\rmn{SFR}=4$ $M_{\odot}/\rmn{yr}$}. See text for further details.
		}
   \label{larsen6946}
\end{figure}
The only difference between the IECMF models that has been applied here are the differing galaxy-wide SFRs. All other parameters are identical to the fit from Fig. \ref{SFRMmaxJ}.
We can also compare the model to the individual galaxies (Figs. \ref{larsen5236} and \ref{larsen6946}) but this has the disadvantage of having higher uncertainties. Nonetheless the model fits quite well, given the uncertainties.\par
\begin{figure}
    \begin{center}
           \resizebox{0.5\textwidth}{!}{\input{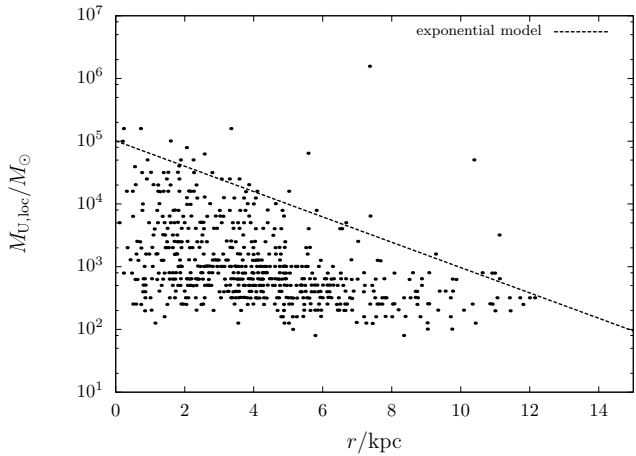}}
    \end{center}
   \caption{
    Comparison of the dependence on the galactocentric distance of $M_{\rmn{U,loc}}$ for the exponential model with observed very young clusters in M33 \citep{jgk}. The model uses \mbox{$\rmn{SFR}=0.16$ $M_{\odot}/\rmn{yr}$}, $\delta t = 10$ Myr and $\beta=2.31$. These values have been chosen so that the model fits the experimental data from the SFR-$M_{\rmn{ecl,max}}$ relation (Fig. \ref{SFRMmaxJ}).
		}
   \label{Mmax-r}
\end{figure}

The $r$-dependency of $M_{\rmn{U,loc}}(r)$ also allows a test of the exponential model. We compare the theoretical $M_{\rmn{U,loc}}$-$r$ dependence with the observed very young star clusters in M33 (data taken from \citealt*{jgk}) in Fig. \ref{Mmax-r}.
Fig. \ref{Mmax-r} shows the $r$-dependence of $M_{\rmn{U,loc}}$ for the model compared to observed very young star clusters.
As $M_{\rmn{U,loc}}$ is the upper mass limit for star clusters at a given galactocentric distance we would expect that no observed cluster is heavier than it. But several observed ones are. Also using the observed SFR (0.16 $M_{\odot}/\rmn{yr}$, \citealt{ski}) in the empirical fit by Weidner (Eq. \ref{maxfit2}) results in a $M_{\rmn{vyc,max}}$ much smaller than several observed clusters. 
But taking into account that individual cluster masses have large uncertainties it becomes apparent that the exponential model fits quite well to  the observed radially decreasing upper masses.

\section{Conclusion}

In this paper we calculated for the first time the galaxy-wide integrated embedded cluster mass function for galaxies, which we assumed for computational ease to be axis-symmetric exponential disks, and showed that it has the form of a Schechter-like function, although locally the ECMF is a pure power-law.\par
To do that we first described an analytical solution of the LECMF. It was assumed there that the theoretical local upper cluster mass limit, $M_{\rmn{U,loc}}(r)$, depends on the galactocentric distance $r$.\par
Integrating the LECMF over the area of the star forming disk yields the galaxy-wide or integrated embedded cluster mass function (IECMF or $\xi_{\rmn{iecl}}$). For this purpose a model describing the $r$-dependence of $M_{\rmn{U,loc}}(r)$ was needed. The exponential model, introduced by \citet{jp}, is found to be in agreement with observational data.\par
Even though the LECMF is a power law, the IECMF resembles a Schechter-like function, the reason being that the upper mass limit of the local power law is defined by $M_{\rmn{U,loc}}(r)$, which decreases with an increasing galactocentric distance.\par
Additionally given a locally estimated power-law index $\beta$ for an ensemble of embedded clusters in a region in a galaxy, we have shown here that other regions elsewhere are expected to have different logarithmic ECMF, depending on the size and position of the region in the galaxy, and that the size of the region implies a Schechter-type turn-down of the embedded cluster mass function. The galaxy-wide ECMF, the integrated ECMF, IECMF, thus becomes a Schechter-type form.\par
All of this opens further possibilities for research. We assumed radially axis-symmetric disk galaxies but it is worth investigating whether a barred galaxy would cause significant changes. Also as has been stated,  the exponential model depends on the theoretical upper limit for the mass of clusters in a galaxy of a specific stellar mass. At the moment we can only give a lower limit to these values, which could underestimate them. Further research in the context of star and cluster formation should be able to give improved insights on the theoretical upper mass limits.
Another step, which has to be done, would be to combine the stellar initial mass function (IMF) and the ECMF into the integrated galactic initial mass function (IGIMF) (\citealt{wei}, \citealt{rec}), which is the galaxy-wide stellar initial mass function.\par

\section*{Acknowledgements}
We thank the referee for helpful comments and S\o ren S. Larsen for providing us with valuable data regarding the galaxies NGC 5236 and NGC 6946. We are also grateful to Jan Pflamm-Altenburg for providing data on M33 and for useful discussions.

\appendix

\section{The LECMF with a Non-Infinitesimal Surface Area}
\label{DeltaA}
In this section we discuss what differences there are to Sec. \ref{LECMF} if one uses an observable non-infinitesimal surface area $\Delta A$ instead of $\mathrm{d}A$. First Eq. (\ref{definition}) becomes

\begin{equation}
	\label{definitionDeltA}
	\xi_{\rmn{lecl}}(M_{\rmn{ecl}}; r)\,\mathrm{d}M_{\rmn{ecl}} \, \Delta A=\mathrm{d}N_{\rmn{ecl}}\, .
\end{equation}

For a given region $\Delta A$, $\Sigma_{\rmn{SFR}}(r)$ becomes the local star formation rate LSFR:

\begin{equation}
	\label{LSFR}
	\rmn{LSFR}(r) = \frac {\Delta \rmn{SFR}}{\Delta A} \, ,
\end{equation}
with $\Delta \rmn{SFR}$ being the star formation rate in the galactic region $\Delta A$.

The total mass formed in stars is then obtained by multiplying $\rmn{LSFR}(r)$ with $\delta t$ (WKL):
\begin{equation}
	\label{Falsch1}
	\rmn{LSFR}(r) \, \delta t= \int_{M_{\rmn{ecl,min}}}^{M_{\rmn{U,loc}}(r)} M_{\rmn{ecl}}^{\prime}\xi_{\rmn{lecl}}(M_{\rmn{ecl}}^{\prime};r)\mathrm{d}M_{\rmn{ecl}}^{\prime}\, .
\end{equation}
For $\beta \neq 2$ we obtain

\begin{equation}
	\label{Kbn2A}
	K(r) = \frac{\rmn{LSFR}(r) \, \delta t (2-\beta)}{M_{\rmn{U,loc}}^{2-\beta}(r)-M_{\rmn{ecl,min}}^{2-\beta}}\, .
\end{equation}
For the special case of $\beta =2$,

\begin{equation}
	\label{Kb2A}
	K(r) = \frac{\rmn{LSFR}(r) \, \delta t }{\ln \left( M_{\rmn{U,loc}}(r)/M_{\rmn{ecl,min}} \right) }\, .
\end{equation}

Using the LECMF it is now possible to calculate the mass of the most massive observable cluster in the region $\Delta A$ ($M_{\rmn{ecl,max,loc}}(r)$). To determine $M_{\rmn{ecl,max,loc}}(r)$ we use two conditions: first there is only one most massive cluster. Second the mass of the most massive cluster is $M_{\rmn{ecl,max,loc}}(r)$.\par
To implement the first condition we choose a mass interval between the upper mass limit $M_{\rmn{U,loc}}(r)$ and $M_{\rmn{ecl,t,loc}}(r)$, with 
$M_{\rmn{ecl,t,loc}}(r)$ chosen in such a way that there is only one cluster between these limits:
\begin{equation}
	\label{Irr1}
	1 \approx \int_{M_{\rmn{ecl,t}}(r)}^{M_{\rmn{U,loc}}(r)} \xi_{\rmn{lecl}}(M_{\rmn{ecl}}^{\prime};r)\mathrm{d}M_{\rmn{ecl}}^{\prime}\, \Delta A \, ,
\end{equation}
and the second condition implies that the mass between these limits is 
$M_{\rmn{ecl,max,loc}}(r)$:
\begin{equation}
	\label{Irr2}
	M_{\rmn{ecl,max,loc}}(r) \approx \int_{M_{\rmn{ecl,t}}(r)}^{M_{\rmn{U,loc}}(r)} M_{\rmn{ecl}}^{\prime} \xi_{\rmn{lecl}}(M_{\rmn{ecl}}^{\prime};r)\mathrm{d}M_{\rmn{ecl}}^{\prime}\, \Delta A\, .
\end{equation}
The reason for the equation being approximated is that $\xi_{\rmn{lecl}}(M_{\rmn{ecl}}^{\prime};r)$ depends on $r$. One would have to perform an integration over $\Delta A$ to get the exact value (see also Sect. \ref{IECMF}). But for small $\Delta A$ the above equation is a good approximation.\par
Using Eqs. (\ref{Irr1}) and (\ref{Irr2}) $M_{\rmn{ecl,max,loc}}(r)$ becomes for $\beta \ne 2$:

\begin{equation}
	\label{Mtn2}
	M_{\rmn{ecl,t}}(r) \approx \left( M_{\rmn{U,loc}}^{1-\beta}(r) - \frac{1-\beta}{K(r)\Delta A} \right)^{\frac{1}{1-\beta}} \, ,
\end{equation}
\begin{equation}
	\label{Mmaxbn2}
	M_{\rmn{ecl,max,loc}}(r) \approx \frac{K(r)}{2-\beta}\left[ M_{\rmn{U,loc}}^{2-\beta}(r)-  M_{\rmn{ecl,t}}^{2-\beta}(r)  \right] \Delta A \, .
\end{equation}
And for $\beta = 2$:

\begin{equation}
	\label{Mmaxb2}
	M_{\rmn{ecl,max,loc}}(r) \approx K(r)\left[ \ln \left( 1 + \frac{M_{\rmn{U,loc}}(r)}{K(r)\Delta A} \right)    \right]\Delta A \, .
\end{equation} 
This is a simple method to determine $M_{\rmn{ecl,max,loc}}(r)$ without having to perform an integration over the area, as long as $\Delta A$ is small enough compared to the galaxy.

\section{Comparison to the WKL normalization}
\label{Weidnernorm}
WKL defined Eq. (\ref{Correct}) slightly differently:
\begin{equation}
	\label{Falsch1}
	\Sigma_{\rmn{SFR}}(r) \, \delta t= \int_{M_{\rmn{ecl,min}}}^{M_{\rmn{ecl,max,loc}}(r)} M_{\rmn{ecl}}^{\prime}\xi_{\rmn{lecl}}(M_{\rmn{ecl}}^{\prime};r)\mathrm{d}M_{\rmn{ecl}}^{\prime}\, .
\end{equation}
Instead of using a theoretical most massive cluster $M_{\rmn{U,loc}}(r)$ as an upper mass limit, $M_{\rmn{ecl,max,loc}}(r)$ was used. This $M_{\rmn{ecl,max,loc}}(r)$ was defined by claiming that there was exactly one cluster in the mass interval $\left[M_{\rmn{ecl,max,loc}}(r), \, M_{\rmn{U,loc}}(r)\right]$:

\begin{equation}
	\label{Falsch2}
	1 \approx \int_{M_{\rmn{ecl,max,loc}}(r)}^{M_{\rmn{U,loc(r)}}} \xi_{\rmn{lecl}}(M_{\rmn{ecl}}^{\prime};r)\mathrm{d}M_{\rmn{ecl}}^{\prime}\, \Delta A \, .
\end{equation}
A criticism of this LECMF formulation is the claim that Eq. (\ref{Falsch2}) would result in one most massive cluster with mass $M_{\rmn{ecl,max,loc}}(r)$. In fact it only ensures that there is a most massive cluster without guaranteeing that this most massive cluster has the mass $M_{\rmn{ecl,max,loc}}(r)$. For this another equation is needed:

\begin{equation}
\label{Falsch4}
  M_{\rmn{ecl,max,loc}}(r) \approx \int_{M_{\rmn{ecl,max,loc}}(r)}^{M_{\rmn{U}}} M_{\rmn{ecl}}^{\prime}\xi_{\rmn{lecl}}(M_{\rmn{ecl}}^{\prime};r)\mathrm{d}M_{\rmn{ecl}}^{\prime}\, \Delta A \, .
\end{equation}
Now it is ensured that there is a mass $M_{\rmn{ecl,max,loc}}(r)$ between $M_{\rmn{ecl,max,loc}}(r)$ and $M_{\rmn{U}}$. But in general an equation system

\begin{equation}
\label{truth}
  \begin{array}{l l}
  1 &= \int_{a}^{b} f(x)\,\mathrm{d}x\, ,\\
  a &= \int_{a}^{b} x f(x)\,\mathrm{d}x\, ,
  
	\end{array}
\end{equation}
does not have a real, non-imaginary, solution for power laws. If $f(x)$ is a distribution function this set of equations requests that the mean of $x$ over the interval $\left[a, \, b\right]$ has the same value as the minimal value of $x$. This can only be possible if \mbox{$a=b$}.\par
The WKL normalization is thus not correct, but the differences obtained when applying it relative to the correct normalization (Eq. \ref{Correct}) are not significant.

\section{Other models}
\label{OTHER}
Alternatives to the exponential model described in Sec. \ref{emj}:
\subsection{Phantom Cluster Model}
\label{opm}
In the past Eq. (\ref{Falsch2}) had been used to calculate the LECMF. It had been assumed that $M_{\rmn{U}}$ is a very large mass, often approximated as infinity. In order to get similar results to the previous method we use a slightly modified form of that equation (introduced in \citealt{sch}):
\begin{equation}
	\label{modFalsch2}
	1 \approx \int_{M_{\rmn{U,loc}}(r)}^{M_{\rmn{\infty}}} \xi_{\rmn{lecl}}(M_{\rmn{ecl}}^{\prime};r)\mathrm{d}M_{\rmn{ecl}}^{\prime}\, \Delta A\, ,
\end{equation}
with $M_{\rmn{\infty}}$ going towards infinity.\par
This resulting model, the phantom cluster model, makes the claim that if there would not be an upper mass limit, then there would be one more cluster in the mass range between the upper mass limit and an infinite mass. A phantom cluster, so to speak.
But there is no physical reason for claiming that there is exactly one cluster between the upper mass limit and infinity \citep{sch}.\par
The resulting relation between $K(r)$ and $M_{\rmn{U,loc}}$ is:
\begin{equation}
	\label{modbn2}
	K(r) \approx \left(\beta-1\right)M_{\rmn{U,loc}}^{\beta-1}(r)/\Delta A\, .
\end{equation}
If one inserts Eq. (\ref{modbn2}) into Eq. (\ref{getKMU}) , so inserting $K(r)$ as a function of $M_{\rmn{U,loc}}(r)$, a direct relation between $M_{\rmn{U,loc}}(r)$ and $r$ is obtained:
\begin{equation}
	\label{getKMUp}
	\int_{M_{\rmn{ecl,min}}}^{M_{\rmn{U,loc}}(r)}\left(\beta-1\right) \frac{M_{\rmn{ecl}}^{1-\beta}}{M_{\rmn{U,loc}}^{1-\beta}(r)\Delta A}\mathrm{d}M_{\rmn{ecl}} \approx\frac{M_{\rmn{tot}}}{2\pi r_{d}^2}e^{-r/r_{d}} \, .
\end{equation}
This equation is not analytically solvable for $M_{\rmn{U,loc}}$. But numerically it is possible to calculate for every $r$ a corresponding $M_{\rmn{U,loc}}(r)$ and $K(r)$. The only remaining free parameters are $\beta$ and $\Delta A$.\par
We can use these two last equations to calculate a $R^{\prime}(M_{\rmn{ecl}})$ for every $M_{\rmn{ecl}}$, as in the exponential model.\par
$\Delta A$ has to be sufficiently small, so that Eq. (\ref{modFalsch2}) is still a good approximation, but has to be large enough to ensure that a complete LECMF can be found in the area. In the following we make the assumption of $\Delta A$ being constant, for simplicity reasons.\par
The free parameters ($\beta$ and $\Delta A$) can be fixed by comparing the model to the empirical SFR-$M_{\rmn{ecl,max}}$ by WKL (see Eq. (\ref{maxfit2}))\par
\begin{figure}
     \begin{center}
            \resizebox{0.5\textwidth}{!}{\input{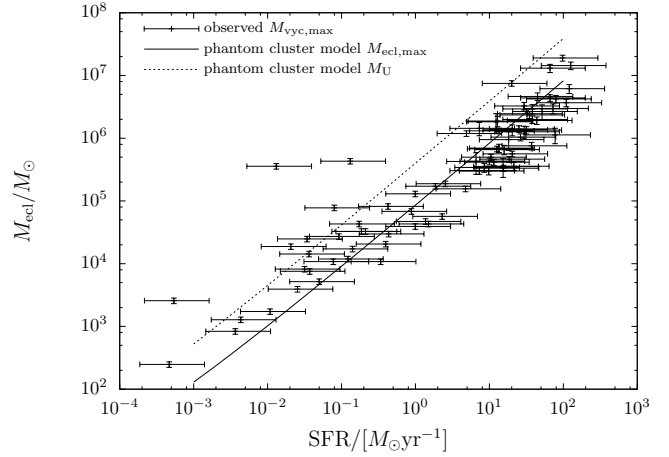}}
    \end{center}
    \caption{
    The observational data are as in Fig. \ref{SFRMmaxJ}. Plotted here is $M_{\rmn{ecl,max}}$ and $M_{\rmn{U}}$ against the SFR as calculated using the phantom cluster model for $\beta=1.73$ and $\Delta A=2.9\, \rmn{kpc}^2$. All observed clusters should be below the $M_{\rmn{U}}$-line within their error-bars, which is not the case here but the data might be consistent with this requirement given the uncertainties. See also \citet{sch} for a discussion of the outlying data points.
		}
   \label{SFRMmaxP}
\end{figure}

Similar to the local case (see Eqs. (\ref{Irr1}) and (\ref{Irr2})) the mass of the heaviest cluster in the entire galaxy ($M_{\rmn{ecl,max}}$) is determined by 
\begin{equation}
	\label{Irr1gP}
	1 = \int_{M_{\rmn{ecl,t}}}^{M_{\rmn{U}}} \xi_{\rmn{iecl}}(M_{\rmn{ecl}}^{\prime})\mathrm{d}M_{\rmn{ecl}}^{\prime}\,
\end{equation}
and

\begin{equation}
	\label{Irr2gP}
	M_{\rmn{ecl,max}} = \int_{M_{\rmn{ecl,t}}}^{M_{\rmn{U}}} M_{\rmn{ecl}}^{\prime} \xi_{\rmn{iecl}}(M_{\rmn{ecl}}^{\prime})\mathrm{d}M_{\rmn{ecl}}^{\prime}\, .
\end{equation}
Using the above Eqs. (\ref{Irr1gP}) and (\ref{Irr2gP}) it is possible to calculate for any $\beta$ a corresponding SFR-$M_{\rmn{ecl,max}}$-relation. These relations can be compared with the empirical found relation from Eq. (\ref{maxfit2}) to constraint $\beta$.
For $\beta=1.73\pm0.01$ and $\Delta A=(2.9\pm0.1)\, \rmn{kpc}^2$ the SFR-$M_{\rmn{ecl,max}}$-curve aligns best to the fit from WKL. $M_{\rmn{U}}$ should be larger than any observed cluster. Given the error bars from the observation this requirement is reasonably fulfilled by the chosen parameters, as can be seen in Fig. \ref{SFRMmaxP}. \par
Another concern is whether the chosen $\Delta A$ is still small enough that Eq. \ref{modFalsch2} is still a good approximation for an integration over the area $\Delta A$. Numerical tests show that using $\Delta A=2.9\, \rmn{kpc}^2$ results in a deviation of up to 3\%, compared to an integration over $\Delta A$, which is still quite accurate. Also observations of the LECMF often use a larger observational area, e.g. \citet{ll} use an area of roughly 12.6 $\rmn{kpc}^2$.

\subsection{Constant $K$ Model}
\label{ckm}
\begin{figure}
    \begin{center}
            \resizebox{0.5\textwidth}{!}{\input{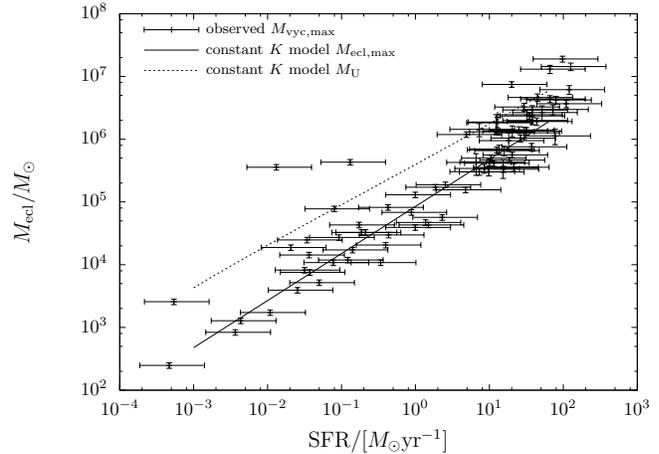}}
    \end{center}
    \caption{
    The observational data are as in Fig. \ref{SFRMmaxJ}. Plotted here is $M_{\rmn{ecl,max}}$ and $M_{\rmn{U}}$, as calculated using the constant $K$ model for $\beta=1.83$ against the SFR. All observed clusters within their uncertainties should be below the $M_{\rmn{U}}$-line, which is fulfilled here well.
		}
   \label{SFRMmaxM}
\end{figure}

The following is an ansatz that tries to be as simple as possible while being consistent with the observational data. If we assume that $K$ does not depend on the galactocentric distance $r$, then Eq. (\ref{getKMU}) should also be valid for \mbox{$r=0$}. Hence $K$ has  for $\beta \neq 2$ the form (with \mbox{$M_{\rmn{U,loc}}(0) =  M_{\rmn{U}}$}):
\begin{equation}
	\label{getKbn2}
	K=\frac{(2-\beta)M_{\rmn{tot}}}{2\pi r_{d}^2\left( M_{\rmn{U}}^{2-\beta}-M_{\rmn{ecl,min}}^{2-\beta}\right)} \, 
\end{equation}
and for $\beta = 2$:

\begin{equation}
	\label{getKb2}
	K=\frac{M_{\rmn{tot}}}{2\pi r_{d}^2\ln \left(\frac{M_{\rmn{U}}}{M_{\rmn{ecl,min}}}\right)} \, .
\end{equation}
The corresponding $M_{\rmn{U,loc}}$ as obtained from Eq. (\ref{getKMU}), using the above form for $K$, is for $\beta \neq 2$:

\begin{equation}
	\label{getMuloc}
	M_{\rmn{U,loc}}(r)=\left[e^{-\frac{r}{r_{d}}} \left(M_{\rmn{U}}^{2-\beta}-M_{\rmn{ecl,min}}^{2-\beta}  \right)  + M_{\rmn{ecl,min}}^{2-\beta} \right]^{\frac{1}{2-\beta}}\, ,
\end{equation}
and for $\beta =2$:

\begin{equation}
	\label{getMuloc2}
	M_{\rmn{U,loc}}(r)=M_{\rmn{ecl,min}} \left(\frac{M_{\rmn{U}}}{M_{\rmn{ecl,min}}}\right)^{e^{-\frac{r}{r_{d}}}} .
\end{equation}
We can use these two last equations to calculate $R^{\prime}(M_{\rmn{ecl}})$, as in the exponential model. For $\beta \neq 2$:
\begin{equation}
	\label{getRbn2}
	R^{\prime}(M_{\rmn{ecl}})=-r_{d}\ln \left[\frac{M_{\rmn{ecl}}^{2-\beta}-M_{\rmn{ecl,min}}^{2-\beta}}{M_{\rmn{U}}^{2-\beta}-M_{\rmn{ecl,min}}^{2-\beta}} \right] \, 
\end{equation}
and for $\beta =2$:

\begin{equation}
	\label{getRb2}
	R^{\prime}(M_{\rmn{ecl}})=-r_{d}\ln \left[\frac{\ln \left( \frac{M_{\rmn{ecl}}}{M_{\rmn{ecl,min}}} \right)}{\ln \left( \frac{M_{\rmn{U}}}{M_{\rmn{ecl,min}}} \right)} \right] \, .
\end{equation}
Consequently $\xi_{\rmn{iecl}}(M_{\rmn{ecl}})$ takes the form
\begin{equation}
	\label{Kintegral}
	\xi_{\rmn{iecl}} (M_{\rmn{ecl}})=K \int_{0}^{2 \pi} \int_{0}^{R^{\prime}(M_{\rmn{ecl}})}M_{\rmn{ecl}}^{-\beta}\,r \, \mathrm{d}r \mathrm{d}\phi\, .
\end{equation}
It is now possible to solve this integral analytically to get an IECMF, which depends only on $M_{\rmn{ecl}}$:

\begin{equation}
	\label{ECMFf}
	\begin{array}{l l}
		\xi_{\rmn{iecl}}(M_{\rmn{ecl}})&=K \pi R^{\prime \, 2}(M_{\rmn{ecl}}) M_{\rmn{ecl}}^{-\beta}\, . \\
	\end{array}
\end{equation}
This model has two remaining free parameters ($\beta$ and $M_{\rmn{U}}$). As in the other models we want to calculate $M_{\rmn{ecl,max}}$ in order to compare it to the empirical SFR-$M_{\rmn{ecl,max}}$ by WKL (see Eq. (\ref{maxfit2})) and constrain these parameters.
Similar to the local case (see Eqs. (\ref{Irr1}) and (\ref{Irr2})) the mass of the heaviest cluster in the entire galaxy ($M_{\rmn{ecl,max}}$) is determined by 
\begin{equation}
	\label{Irr1g}
	1 = \int_{M_{\rmn{ecl,t}}}^{M_{\rmn{U}}} \xi_{\rmn{iecl}}(M_{\rmn{ecl}}^{\prime})\mathrm{d}M_{\rmn{ecl}}^{\prime}\,
\end{equation}
and

\begin{equation}
	\label{Irr2g}
	M_{\rmn{ecl,max}} = \int_{M_{\rmn{ecl,t}}}^{M_{\rmn{U}}} M_{\rmn{ecl}}^{\prime} \xi_{\rmn{iecl}}(M_{\rmn{ecl}}^{\prime})\mathrm{d}M_{\rmn{ecl}}^{\prime}\, .
\end{equation}
Therefore we can find for every $\beta$ a SFR-$M_{\rmn{ecl,max}}$-curve that aligns with Eq. (\ref{maxfit2}). Doing so we can find for every $\beta$ a corresponding $M_{\rmn{U}}$:
\begin{equation}
	\label{MUbeta}
	M_{\rmn{U}} = \frac{74138}{2-\beta}\rmn{SFR}^{0.91-0.15 \beta}\, .
\end{equation}
In order to be realistic this $M_{\rmn{U}}$ has to be heavier than any relevant observed cluster for this specific SFR, but also not so much larger than the mass of the heaviest observed cluster that there would be an unrealistic gap between them. As can be seen in Fig. \ref{SFRMmaxM}, $\beta=1.83\pm0.1$ fulfils these observational constraints well.

\subsection{Comparison of the three models}
\begin{figure}
    \begin{center}
            \resizebox{0.5\textwidth}{!}{\input{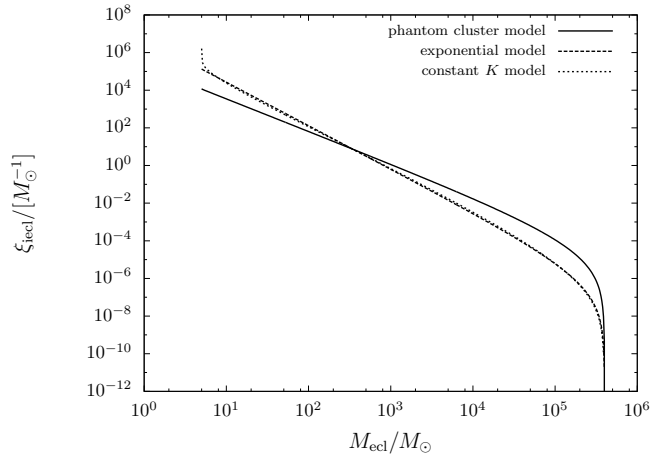}}
    \end{center}
    \caption{
    Comparison of the three models for a galaxy-wide SFR of 1 $M_{\odot}/\rmn{yr}$ and for $\delta t=10$ Myr. For the phantom cluster model $\beta=1.73$ was used, for the exponential model $\beta=2.31$ and for the constant $K$ model $\beta=1.83$. These values have been chosen so that the models fit the observational data shown in Figs. \ref{SFRMmaxJ}, \ref{SFRMmaxP} and \ref{SFRMmaxM}.
		}
   \label{ECMF_P}
\end{figure}

All the models can be written as 
\begin{equation}
	\label{generalmodel}
	\xi_{\rmn{iecl}}(M_{\rmn{ecl}})=f(M_{\rmn{ecl}}, \beta) M_{\rmn{ecl}}^{-\beta}\, ,
\end{equation}
with $f(M_{\rmn{ecl}}, \beta)$ varying from model to model. 
All three models use a different $f(M_{\rmn{ecl}}, \beta)$ and also different values for $\beta$ in order to be in agreement to the empirical fit by WKL (Eq. \ref{maxfit2}). Eq. (\ref{generalmodel}) shows that $\xi_{\rmn{iecl}}$ is not a pure power-law, so $\beta$ is not the logarithmic slope of $\xi_{\rmn{iecl}}$. For a comparison of ECMFs resulting from the best fits of the models, see Fig. \ref{ECMF_P}.
\begin{figure}
    \begin{center}
            \resizebox{0.5\textwidth}{!}{\input{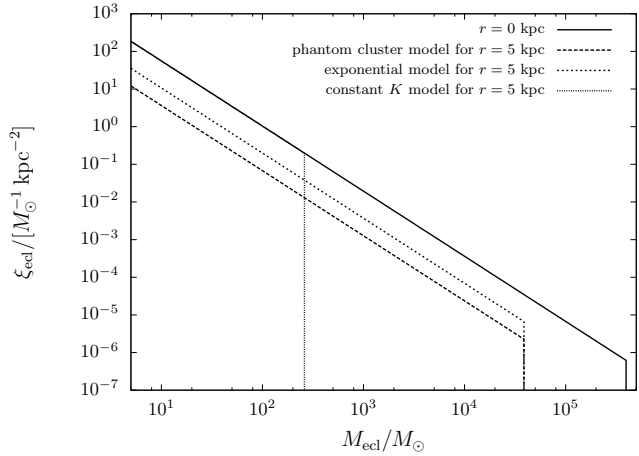}}
    \end{center}
    \caption{
    Comparison of the dependence on the galactocentric distance of the three models. All models are shown with $\beta=1.73$, $M_{\rmn{U}}=400000$ $M_{\odot}$, $\rmn{SFR}=1$ $M_{\odot}/\rmn{yr}$, $\delta t=10$ Myr and $r_{d}=2.15$ kpc. These values are not the previous fit values but were chosen so that all models produce the same LECMF for a galactocentric distance of 0 kpc. Note that the constant $K$ model leads to a significantly smaller $M_{\rmn{U,loc}}(r)$, for \mbox{$r=5$ kpc}, than the other models.
		}
   \label{LECMF_compare}
\end{figure}
One can see that the exponential (Sec. \ref{emj}) and the constant $K$ (Sec. \ref{ckm}) model look very similar, although they have different values for $\beta$. This is because in these models $\beta$ is no longer the logarithmic slope. In contrast to the exponential and the constant $K$ model, the phantom cluster model looks different: its logarithmic slope is less steep.\par
The models also result in different dependencies on the distance to the the galactic center in the case of the LECMF. This is illustrated in Fig. \ref{LECMF_compare} with parameters chosen in such a way that the models result in the same LECMF for $r=0$ kpc.
\par
An important difference of the LECMF between the phantom cluster model (Sec. \ref{opm}) and the other models is that the phantom cluster model has the free parameters $\beta$ and $\Delta A$, in comparison to the other two which have instead of $\Delta A$ a direct dependence on the parameter $M_{\rmn{U}}$.
Therefore the exponential and the constant $K$ model allow for a $M_{\rmn{U}}$ which is larger than the masses of the observed clusters (see Figs. \ref{SFRMmaxJ} and \ref{SFRMmaxM}).\par
Another difference is that the constant $K$ model can be solved analytically, whereas the other models need to be solved numerically.\par
Summarizing, overall the exponential model, which is physically motivated \citep{jp}, works best, as the phantom cluster model depends on a rather arbitrary $\Delta A$ and the constant $K$ model does not reproduce well local data. But one may still apply these models, e.g. if one needs an IECMF without numerical modelling, one can use the constant $K$ model.

\section{Summary of Variables}
\begin{table*}

\caption{List of variables that were introduced.}
\label{mathmode}
\begin{tabular}{@{}llllll}
Variable & Definition & Reason for Introduction\\
\hline
$N_{\rmn{ecl}}$     & number of embedded star clusters       & see definition of $\xi_{\rmn{lecl}}$     \\[2pt]
$M_{\rmn{ecl}}$ & mass of an embedded star cluster         & see definition of $\xi_{\rmn{lecl}}$  \\[2pt]
$A$    & surface area        & see definition of $\xi_{\rmn{lecl}}$     \\[2pt]
$r$    & galactocentric distance    & see definition of $\xi_{\rmn{lecl}}$   \\[2pt]
$\xi_{\rmn{lecl}}$    & local embedded cluster mass function: $\frac{\mathrm{d}N}{\mathrm{d}M_{\rmn{ecl}} \mathrm{d} A} $ at $r$     & parameter studied in this paper   \\[2pt]
$K$  & normalization constant    & the normalization constant for $\xi_{\rmn{lecl}}$  \\[2pt]
$\beta$    & power law index     & power law index and logarithmic slope of $\xi_{\rmn{lecl}}$   \\
$M_{\rmn{ecl,min}}$ & minimal embedded cluster mass & the smallest cluster mass \\
$M_{\rmn{ecl,max}}$ & the largest cluster mass in a given galaxy & used to compare theory with observations  \\
$M_{\rmn{ecl,max,loc}}$ & local $M_{\rmn{ecl,max}}$ & used to compare theory with observations  \\
$\delta t$ & star formation time-scale & proportional to $K$ \\
SFR & star formation rate & used to calculate the normalization constant $K$ \\
LSFR & local star formation rate & proportional to $K$ in the local case \\
$\Sigma_{\rmn{SFR}}$ & proportional to $K$ in the local, infinitesimal case \\
$M_{\rmn{U}}$ & theoretical most massive cluster physically possible & upper limit for $M_{\rmn{ecl}}$ in a galaxy  \\
$M_{\rmn{U,loc}}$ & local $M_{\rmn{U}}$ & upper limit for $M_{\rmn{ecl}}$ at $r$  \\
$M_{\rmn{ecl,t}}$ & auxiliary variable & needed together with $M_{\rmn{U,loc}}$ to calculate $M_{\rmn{ecl,max,loc}}$:  \\
& &  there is  exactly one cluster of mass $M_{\rmn{ecl,max,loc}}$\\
& & between $M_{\rmn{ecl,t}}$ and $M_{\rmn{U,loc}}$\\
$\xi_{\rmn{iecl}}$    & integrated embedded cluster mass function : $\frac{\mathrm{d}N}{\mathrm{d}M_{\rmn{ecl}}} $      & parameter studied in this paper \\[2pt]
$R_{\rmn{gal}}$    & radius of the star forming area& upper limit for $r$ \\[2pt]
$R^{\prime}$    & maximal theoretical galactocentric distance for a cluster & necessary to change the $r$-dependence into a mass dependence \\[2pt]
$r_{d}$    & disk scale length & necessary to describe the exponential galactic disk \\[2pt]
$M_{\rmn{vyc,max}}$ & observationally derived maximal very young cluster mass & a good approximation for $M_{\rmn{ecl,max}}$  \\
WKL & \citet*{wkl} & gets cited often in this paper  \\
\hline
\end{tabular}

\end{table*}

%%%%%%%%%%%%%%%%%%%%%%%%%%%%%%%%%%%%

\makeatletter
% define \thebiblio (same as thebibliography, but
% without the section heading)
\def\thebiblio#1{%
 \list{}{\usecounter{dummy}%
         \labelwidth\z@
         \leftmargin 1.5em
         \itemsep \z@
         \itemindent-\leftmargin}
 \reset@font\small
 \parindent\z@
 \parskip\z@ plus .1pt\relax
 \def\newblock{\hskip .11em plus .33em minus .07em}
 \sloppy\clubpenalty4000\widowpenalty4000
 \sfcode`\.=1000\relax
}
\let\endthebiblio=\endlist
\makeatother

%%%%%%%%%%%%%%%%%%%%%%%%%%%%%%%%%%%%

% \bsp % ``This paper has been produced using the ...''

\label{lastpage}


\begin{thebibliography}{}
   \bibitem[\protect\citeauthoryear{Baumgardt \& Makino}{2003}]{baum} Baumgardt H., Makino J., 2003, MNRAS, 340, 227
   \bibitem[\protect\citeauthoryear{Bovy \& Rix}{2013}]{bovy} Bovy J., Rix H-W., 2013, ApJ, 779, 115
   \bibitem[\protect\citeauthoryear{Burkert \& Smith}{2000}]{bur} Burkert A., Smith G. H., 2000, ApJ, 542, L95
  \bibitem[\protect\citeauthoryear{Dopita et al.}{2010}]{dop} Dopita M. A. et al.,  2010,
    ApJ, 710, 964
   \bibitem[\protect\citeauthoryear{Egusa, Sofue \& Nakanishi}{Egusa et al.}{2004}]{egu} Egusa F., Sofue Y., Nakanishi H., 2004, PASJ, 56, L45
      \bibitem[\protect\citeauthoryear{Fukui et al.}{1999}]{fuk} Fukui Y. et al., 1999, PASJ, 51, 745
  \bibitem[\protect\citeauthoryear{Gieles et al.}{2006}]{gl} Gieles M., Larsen S. S., Bastian N., Stein I. T., 2006a, A\&A, 450, 129
  \bibitem[\protect\citeauthoryear{Heesen et al.}{2014}]{hee} Heesen V., Brinks E., Leroy A. K., Heald G., Braun R., Bigiel F., Beck R., 2014, AJ, 147, 103
  \bibitem[\protect\citeauthoryear{Hong et al.}{2013}]{hong} Hong S., Calzetti D., Gallagher III J. S., Martin C. L., Conselice C. J., Pellerin A., 2013, ApJ, 777, 63
    \bibitem[\protect\citeauthoryear{Jordan et al.}{2007}]{jord} Jordan A. et al., 2007, ApJS, 171, 101
        \bibitem[\protect\citeauthoryear{Kirk \& Myers}{2012}]{km} Kirk H., Myers P. C., 2012,
  ApJ, 745, 131
          \bibitem[\protect\citeauthoryear{Kroupa}{2005}]{pakr} Kroupa P., 2005,
  ESA Special Publ., 576, 629
          \bibitem[\protect\citeauthoryear{Kroupa \& Bouvier}{2003}]{kb} Kroupa P., Bouvier J., 2003,
  MNRAS, 346, 343
  \bibitem[\protect\citeauthoryear{Kroupa et al.}{2013}]{pa} Kroupa P., Weidner C., Pflamm-Altenburg J., Thies I., Dabringhausen J., Marks M., Maschberger T., 2013,
    Planets, Stars and Stellar Systems. Volume 5: Galactic Structure and Stellar
Populations, Springer Science+Business Media Dordrecht
      \bibitem[\protect\citeauthoryear{Lada \& Lada}{2003}]{ll} Lada C. J., Lada E. A., 2003,
  ARA\&A, 41, 57
        \bibitem[\protect\citeauthoryear{Larsen}{1999}]{lars} Larsen S. S., 1999,
  A\&AS, 139, 393
        \bibitem[\protect\citeauthoryear{Larsen}{2009}]{lars2} Larsen S. S., 2009,
   A\&A, 494, 539
  \bibitem[\protect\citeauthoryear{Megeath et al.}{2016}]{meg} Megeath S.T. et al.,  2016,
    AJ, 151, 5
      \bibitem[\protect\citeauthoryear{Meingast et al.}{2016}]{mei} Meingast S. et al, 2016,
    A\&A, 587, 153
       \bibitem[\protect\citeauthoryear{Parmentier \& Gilmore}{2007}]{par} Parmentier G., Gilmore, G., 2007, MNRAS, 377, 352
  \bibitem[\protect\citeauthoryear{Pflamm-Altenburg, Gonz\'alez-L\'opezlira \& Kroupa}{Pflamm-Altenburg et al.}{2013}]{jgk} Pflamm-Altenburg J., Gonz\'alez-L\'opezlira R. A., Kroupa P., 2013,
    MNRAS, 435, 2604
  \bibitem[\protect\citeauthoryear{Pflamm-Altenburg \& Kroupa}{2008}]{jp} Pflamm-Altenburg J., Kroupa P., 2008,
    Nature, 455, 641
\bibitem[\protect\citeauthoryear{Porcel et al.}{1998}]{Por} Porcel C., Garzon F., Jimenez-Vicente J., Battaner E., 1998, A\&A, 330, 136
\bibitem[\protect\citeauthoryear{Randriamanakoto et al.}{2013}]{rev} Randriamanakoto Z., Escala A., V\"ais\"anen P., Kankare E., Kotilainen J., Mattila S., Ryder S., 2013, ApJ, 775, L38
\bibitem[\protect\citeauthoryear{Recchi \& Kroupa}{2015}]{rec} Recchi S., Kroupa P., 2015, MNRAS, 446, 4168
\bibitem[\protect\citeauthoryear{Schulz, Pflamm-Altenburg \& Kroupa}{Schulz et al.}{2015}]{sch} Schulz C., Pflamm-Altenburg J., Kroupa P., 2015, A\&A, 582, 93
\bibitem[\protect\citeauthoryear{Skibba et al.}{2011}]{ski} Skibba R. A. et al., 2011, ApJ, 738, 89
  \bibitem[\protect\citeauthoryear{Tamburro et al.}{2008}]{tam} Tamburro D., Rix H. W., Walter F., Brinks E., de Blok W. J. G., Kennicutt R. C., Mac Low M. M., 2008,
    AJ, 136, 2872
  \bibitem[\protect\citeauthoryear{Weidner, Kroupa \& Larsen}{Weidner et al.}{2004}]{wkl} Weidner C., Kroupa P., Larsen S. S., 2004,
    MNRAS, 350, 1503
  \bibitem[\protect\citeauthoryear{Weidner et al.}{2013}]{wei} Weidner C., Kroupa P., Pflamm-Altenburg J., Vazdekis A., 2013,
    MNRAS, 436, 3309
      \bibitem[\protect\citeauthoryear{Yamaguchi et al.}{2001}]{yam} Yamaguchi R. et al., 2001,
    PASJ, 53, 985

\end{thebibliography}
\end{document}